\documentclass{iaus}
\usepackage{graphicx}

\title[Asteroseismology of close binary stars]
{Asteroseismology of close binary stars}

\author[Conny Aerts]   
{Conny Aerts$^{1,2}$}

\affiliation{$^1$Instituut voor Sterrenkunde, Celestijnenlaan 200D, B-3001
  Leuven, Belgium\break email: conny@ster.kuleuven.be\\[\affilskip]
$^2$ Department of Astrophysics, Radboud University Nijmegen, P.O.Box 
 9010, 6500 GL Nijmegen, The Netherlands}

\pubyear{2007}
\volume{240}  
\pagerange{1--10}
\date{?? and in revised form ??}
\jname{Binary Stars as Critical Tools \& Tests
in \\Contemporary Astrophysics}
\editors{W.\ Hartkopf, E.\ Guinan, and P.\ Harmanec, eds.}
\begin{document}

\maketitle

\begin{abstract}
In this review paper, we summarise the goals of asteroseismic studies
  of close binary stars. We first briefly recall the basic principles of
  asteroseismology, and highlight how the binarity of a star can be an asset,
  but also a complication, for the interpretation of the stellar
  oscillations. We discuss a few sample studies of pulsations in close binaries
  and summarise some case studies. This leads us to conclude that
  asteroseismology of close binaries is a challenging field of research, but
  with large potential for the improvement of current stellar structure
  theory. Finally, we highlight the best observing strategy to make efficient
  progress in the near future.
\keywords{
(stars:) binaries: general,
(stars:) binaries (including multiple): close,
(stars:) binaries: eclipsing,
stars: early-type,
stars: evolution,
stars: interiors,
stars: oscillations (including pulsations),
(stars:) subdwarfs,
stars: variables: other,
stars: statistics}
\end{abstract}

\firstsection 

\section{Goals and current status of asteroseismology}

The main goal of asteroseismology is to improve the input physics of stellar
structure and evolution models by requiring such models to fit observed
oscillation frequencies. The latter are a very direct and high-precision probe
of the stellar interior, which is otherwise impossible to access. In particular,
asteroseismology holds the potential to provide a very accurate stellar age
estimate from the properties of the oscillations near the stellar core, whose
composition, stratification and extent is captured by the frequency behaviour.
In general, the oscillations are characterised by their frequency $\nu$ and
their three wavenumbers $(\ell,m,n)$ determining the shape of the eigenfunction
(e.g.\ Unno et al.\ 1989). Seismic tuning of interior stellar structure becomes
within reach when a unique set of values $(\nu,\ell,m,n)$ is assigned to each of
the observed oscillation modes. In practice, the number of identified modes
needed to improve current structure models depends on the kind of star.  For
B-type stars on the main sequence, e.g., even two or three well-identified modes
can sometimes be sufficient to put constraints on the internal rotation profile
(e.g.\ Aerts et al.\ 2003, Pamyatnykh et al.\ 2004) and/or on the extent of the
convective core (e.g.\ Aerts et al.\ 2006, Mazumdar et al.\ 2006).

Asteroseismology received a large impetus after the very successful application
of the technique to the Sun during the past decade. Helioseismology indeed
revolutionised our understanding of the solar structure, including the solar
interior rotation and mixing (e.g.\ Christensen-Dalsgaard 2002 for an extensive
review).  This remains true today, even though helioseismology is presently
undergoing somewhat of a crisis with the revision of the solar abundances
(Asplund et al.\ 2005, and references therein). These led to a slightly
diminished precision, but still the relative agreement between the observed
solar oscillation properties and those derived from the best solar structure
models remains below 0.5\% for most of the basic quantities, such as the
interior sound speed and composition profiles (while it used to be below 0.1\%
with the old solar abundances). Helioseismology thus provided us with a {\it
unique\/} calibrator to study the structure of other stars.

However, the Sun is just one simple star. It is a slow rotator, it is hardly
evolved, it does not possess a convective core, it does not suffer from severe
mass loss, etc. There are thus a number of effects, of great importance for
stellar evolution, that have not yet been tested with high accuracy. Since stars
of different mass and evolutionary stage have very different structure, we
cannot simply extrapolate the solar properties across the whole HR
diagram. Stellar oscillations allow us to evaluate our assumptions on the input
physics of evolutionary models for stars in which these effects are of
appreciable importance. For the moment, such oscillations are the only accurate 
probe of the interior physics that we have available.

At present, the interior mixing processes in stars are often described by
parametrised laws, such as the time-independent mixing length formulation for
convection. Values near the solar ones for the mixing length and the convective
overshoot are often assumed, by lack of better information.  Similarly, rotation
is either not included or with an assumed rotation law in stellar models, while
an accurate description of rotational mixing is crucial for the evolution of
massive stars (e.g.\ Maeder \& Meynet 2000).

Fortunately, both core overshoot and rotation modify the frequencies of the
star's oscillations and they do it in a different way.  Core overshoot values
can be derived from zonal oscillation modes, which have $m=0$, because they are
not affected by the rotation of the star for cases where the centrifugal force
can be ignored, but they are strongly affected if an overshoot region surrounds
the well-mixed convective core. On the other hand, the rotational splitting of
the oscillation frequencies is dependent of the internal rotation profile. This
can be mapped from the identification of the modes with $m\neq 0$ once the
central $m=0$ component of these modes has been fixed by the models.  Adequate
seismic modelling of core convection and interior rotation is thus within reach,
provided that one succeeds in the identification of at least two, and preferably
a much larger set of $(\nu,\ell,m,n)$. This observational requirement demands
combined high-precision multicolour photometric and high-resolution
spectroscopic measurements with a high duty cycle (typically above 50\%).  For
an extensive introduction into asteroseismology, its recent successes, and its
challenges, I advise the review papers by Kurtz (2006) and by De Ridder
(2006). None of the successful cases so far concerns a close binary \ldots

\section{The specific case of oscillations in close binaries}

Close binary stars have always played a crucial role in astrophysics, not only
because, besides pulsating stars, they allow stringent tests of stellar
evolution models, but also because they are laboratories in which
specific physical processes, which do not occur in single stars, take place.
Understanding these processes is important because at least half of all stars
occur in multiple systems.  

Close binaries are subjected to tidal forces and can evolve quite differently
than single stars.  Seismic mass and age estimates of pulsating components in
close binaries of different stages of evolution, would allow to refine the
binary scenarios in terms of energy loss and to probe the interior structure of
stars subject to tidal effects in terms of angular momentum transport through
non-rigid internal rotation.

\subsection{Overview of observational data}

Two excellent review papers on pulsating stars in binaries and multiple systems
(including clusters) are available in Pigulski (2006) and Lampens (2006). These
are highly recommended to the reader who wants to get a clear overview of the
observational status and become familiar with this subfield of binary star
research. One learns from these works that numerous pulsating stars are known in
binaries, that lots of open questions remain concerning the confrontation
between tidal theory and observational data, and that the best cases to monitor
in the future are pulsating stars in eclipsing binaries.  Eclipsing binaries
have indeed revealed values for the core extent in B stars in excess of those
found from asteroseismology of single B stars (Guinan et al.\ 2000). A natural
thing to do would be to repeat the type of asteroseismic studies that led to the
core overshoot value and internal rotation profile of single stars, as discussed
in the previous section, but then for pulsating stars in eclipsing
binaries. This would allow to disentangle the core overshoot from the internal
rotation with higher confidence level than for single stars. In this respect, I
refer to Pigulski et al., Golovin \& Pavlenko and Latkovic (these proceedings)
for new discoveries of pulsating stars in eclipsing binaries and to Br\"untt et
al.\ (these proceedings) for the best quality data available of such systems to
date.

\subsection{Mode identification through eclipse mapping}

A remark worth giving here is the potential to perform mode identification
through the technique of eclipse mapping in eclipsing binaries with a pulsating
component. This idea was put forward more than 30 years ago by Nather \&
Robinson (1974) who interpreted the phase jumps of 360$^\circ$ in the nova-like
binary UX\,UMa in terms of non-radial oscillation modes of $\ell=2$. We now know
that this interpretation was premature and that the observed phase phenomenon is
far better explained in terms of an oblique rotator model.  

Mkrtichian et al.\ (2004) excluded odd $\ell+m$ combinations for the Algol-type
eclipsing binary star AS\,Eri from the fact that the disk-integrated amplitude
disappears during the eclipse.  Gamarova et al.\ (2005) and Rodr\'{\i}guez et
al.\ (2004) made estimates of the wavenumbers for the Algol-type eclipsing
binaries AB\,Cas and found a dominant radial mode, in agreement with the
out-of-eclipse identification.  By far the best documented version of mode
identification from photometric data using eclipse mapping is available in Reed
et al.\ (2005). While their primary goal was to search for evidence of tidally
tipped pulsation axes in close binaries, they also made extensive simulations,
albeit for the very specific case of eclipse mapping of pulsating subdwarf B
star binaries.  They find that $\ell>2$ modes become visible during an eclipse
while essentially absent outside of eclipse. Their tools have so far only been
applied to the concrete cases of KPD\,1930+2752 and of PG\,1336-018 (Reed et
al.\ 2006) but without clear results.

We must conclude that, still today, more than 30 years after the original idea,
mode identification from eclipse mapping is hardly applied successfully in
practice, and it certainly has not been able to provide constraints on the
wavenumbers for stars which have been modelled seismically. New promising work
along this path is, however, in progress (Mkrtichian, private communication).

\subsection{Pressure versus gravity modes}

One thing to keep in mind is that there are two types of oscillations from the
viewpoint of the acting forces, and that only one type is relevant in the
context of tidal excitation. One either has pressure modes, for which the
dominant restoring force is the pressure, or gravity modes, for which the
dominant restoring force is buoyancy. Tidal excitation can only occur whenever
the orbital frequency is an integer multiple of the pulsation frequency, the
integer typically being smaller than ten. This follows from the expression of
the tide-generating gravitational potential (e.g.\ Claret et al.\ 2005, and
references therein). Moreover, in that case, one expects only $\ell =2$ modes to
be excited, with an $m$-value that provides the good combination between the
rotation, oscillation and orbital frequencies to achieve a non-linear
resonance. Such a situation is much more likely to occur for gravity modes,
which have, in main-sequence stars, oscillation periods of order days, than for
pressure modes which have much shorter periods of hours. The same holds true for
compact oscillators, whose pressure modes have short periods of minutes while
their gravity modes have periodicities of hours and these may be of the same
order as the orbital periods.

Tidal forces can, of course, alter the free oscillations excited in components
of binaries. In that case, one expects to see shifts of the frequencies of the
free oscillation modes with values that have something to do with the orbital
frequency. This alteration can, but does not need to be, accompanied with
ellipsoidal variability.  In the latter case, the tides have deformed the
oscillator from spherical symmetry, and one needs to take into account this
deformation in the interpretation of the oscillation modes.

\subsection{Sample studies}

Soydugan et al.\ (2006) have presented a sample study of 20 eclipsing binaries
with a $\delta\,$Sct-type component. They came up with a linear relation between
the pulsation and orbital period:
\begin{equation}
P_{\rm puls}\ =\ (0.020\pm 0.002)\ P_{\rm orb}\ -\ (0.005\pm 0.008).
\end{equation}
This observational result implies that tidal excitation cannot be active in this
sample of binaries, because this would demand a coefficient larger than
typically 0.1 as mentioned above, i.e.\ an order of magnitude larger than the
observed one.  

A similar conclusion was reached by Fontaine et al.\ (2003), who investigated if
the oscillations in pulsating subdwarf B stars could be tidally induced, given
that 2/3 of such stars are in close binaries. Tidal excitation can, at most,
explain some of the gravity modes observed in some such stars but this is not
yet proven observationally.  On the other hand, all the formation channels for
subdwarf B stars involve close binary evolution (Han et al., Pulstylnik \&
Pustynski, Morales-Rueda et al., these proceedings).  An asteroseismic
high-precision mass and age estimate of such a star would imply stringent
constraints on the proposed scenarios and on the role of the binarity for the
oscillatory behaviour (see Hu et al., these proceedings).

Aerts \& Harmanec (2004), finally, made a compilation of some 50 confirmed
line-profile variables in close binaries (mainly OBA-type stars). They could not
find any significant relation between the binary and variability parameters of
these stars.

We come to the important conclusion that we have by no means a good statistical
understanding of the effects of binarity on the components' oscillations. This
situation can only be remedied by performing several case studies
of pulsating close binaries in much more depth than those existing at present.

\section{Towards successful seismic modelling of binaries}

In general, we have to make a distinction between three different situations
when studying oscillations in close binaries with the goal to make seismic
inferences of the stellar structure.

\subsection{Reduction of the error box of fundamental parameters}

In a first case, the binarity is simply an asset for the asteroseismologist,
because it allows for a reduction of the observational error box of the
fundamental parameters of the pulsating star. This case is relevant whenever we
lack an accurate parallax value, and thus a good estimate of the luminosity and
the mass. This is mainly the case for OB-type stars, but sometimes also for
cooler stars or compact objects. In absence of a good mass or luminosity
estimate, the asteroseismologist cannot discriminate sufficiently between the
seismic models fulfilling the observed and identified oscillation modes.  It was
recently shown that a combined observational effort based on interferometry and
high-precision spectroscopy can enhance significantly the precision of
luminosity and mass estimates in double-lined spectroscopic binaries with a
pulsating component (Davis et al., these proceedings), if need be after
spectroscopic disentangling as in Ausseloos et al.\ (2006).  Such spectroscopy
is in any case also needed to derive the oscillation wavenumbers $(\ell,m)$
(e.g.\ Briquet \& Aerts 2003, Zima 2006).

The best case study of a pulsating star whose binarity was of great help in the
seismic interpretation is the one of $\alpha\,$CenA+B with two pulsating
components (Miglio \& Montalban 2006, and references therein). The binarity gave
such stringent constraints in this case, that in-depth information on the
interior of both components was found. In particular, it was found that the
components seem to have different values of the convective overshooting
parameter and that the primary, being of the same spectral type as the Sun, is
right at the limit of having or not a small convective core. The seismic
analysis also provided an accurate age estimate of the system.  We refer to
Miglio \& Montalban (2006), and references therein, for the latest results and
an overview of the stellar modelling.

Other well-known, but less successful examples are the $\delta\,$Sct star
$\theta^2\,$Tau (Breger et al.\ 2002, Lampens et al.\ these proceedings) and the
B-type pulsators $\beta\,$Cen (Ausseloos et al.\ 2006), $\lambda\,$Sco (Tango et
al.\ 2006) and $\psi\,$Cen (Br\"untt et al.\ 2006). For these four stars, which
are all fairly rapid rotators, our comprehension of the observed pulsational
behaviour is at present insufficient for detailed seismic inference of their
interior structure.  In particular, we lack reliable mode identification of the
detected frequencies. Not being able to identify the oscillation modes properly
is the largest stumbling block in asteroseismology of single stars as well.

\subsection{Tidal perturbations of free oscillations}

The second case concerns seismic targets whose free oscillation spectrum is
affected by the tides. The first such situation was reported for the
$\delta\,$Sct star 14\,Aur\,A, a 3.8d circular binary whose close frequency
splitting of an $\ell =1$ mode does not match the one of a single rotating star
and was interpreted in terms of a tidal effect by Fitch \& Wisniewski (1979).
Alterations of the free oscillation modes by tidal effects have also been
claimed for the three $\beta\,$Cep stars $\alpha\,$Vir (also named Spica) which
has an orbital period $P_{\rm orb}=4.1$d and an eccentricity $e=0.16$
(Aufdenberg, these proceedings, Smith 1985a,b), $\sigma\,$Sco with $P_{\rm
orb}=33$d, $e=0.44$ (Goossens et al.\ 1984) and $\eta\,$OriAab with $P_{\rm
orb}=8$d, $e=0.01$ (De Mey et al.\ 1996).

When, besides oscillation frequencies, the orbital frequency and its harmonic is
found in the frequency spectrum, one is dealing with oscillations in an
ellipsoidal variable. This case occurs for the $\beta\,$Cep stars $\psi^2\,$Ori
with $P_{\rm orb}=2.5$d, $e=0.05$ (Telting et al.\ 2001) and $\nu\,$Cen with
$P_{\rm orb}=2.6$d, $e=0$ (Schrijvers \& Telting 2002), as well as for the
$\delta\,$Sct stars $\theta\,$Tuc with $P_{\rm orb}=7$d, $e=0$ (De Mey et al.\
1998), XX\,Pyx with $P_{\rm orb}=1.2$d, $e=0$ (Aerts et al.\ 2002) and
HD\,207251 with $P_{\rm orb}=1.5$d, $e=0$ (Henry et al.\ 2004). The deformation
of the pulsator has so far been neglected in the interpretation of the
oscillation frequencies in these binaries. The reason is clear: the complexity
of the mathematical description of non-radial oscillations for a deformed star
is huge compared to the case of a spherically-symmetric star.  Aerts et al.\
(2002) have pointed out that this omission may well be the reason why the
extensive efforts to model XX\,Pyx seismically, as in Pamyatnykh et al.\ (1998),
have failed so far.
\begin{figure}
\includegraphics[height=4.5in,width=5.5in,angle=0]{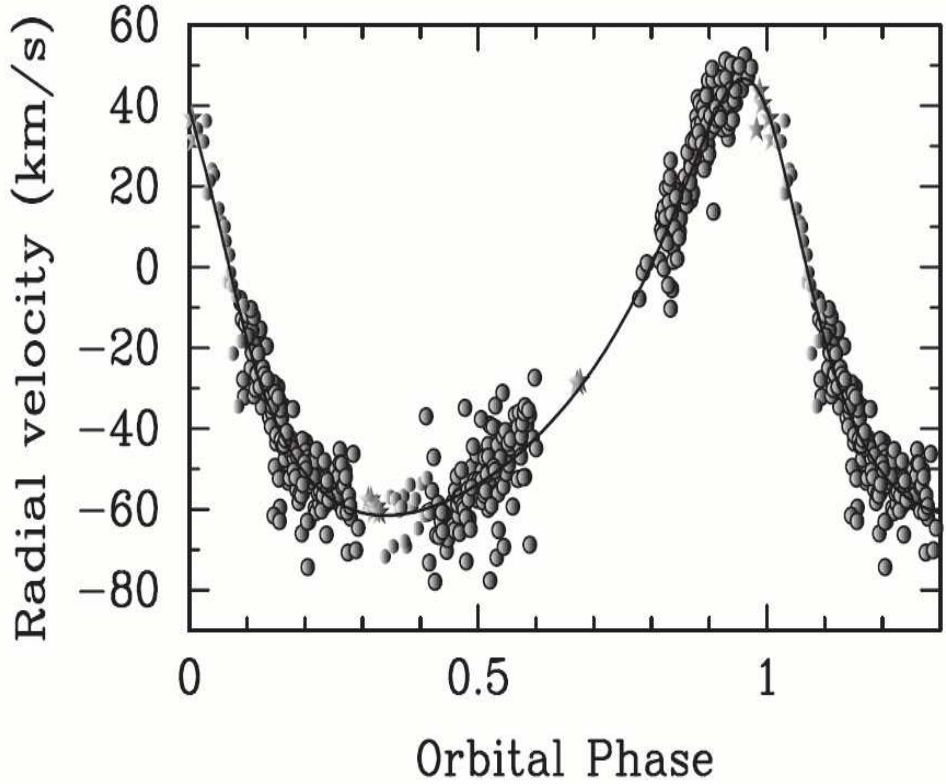}
\caption{Observed radial velocities of HD\,209295 (symbols) phased with the
orbital solution (full line). For an explanation of the symbols, we refer to 
Handler et al.\ (2002), from which this figure was reproduced with permission
from MNRAS.} 
\label{orbit}
\end{figure}

\begin{figure}
\includegraphics[height=5in,width=3in,angle=0]{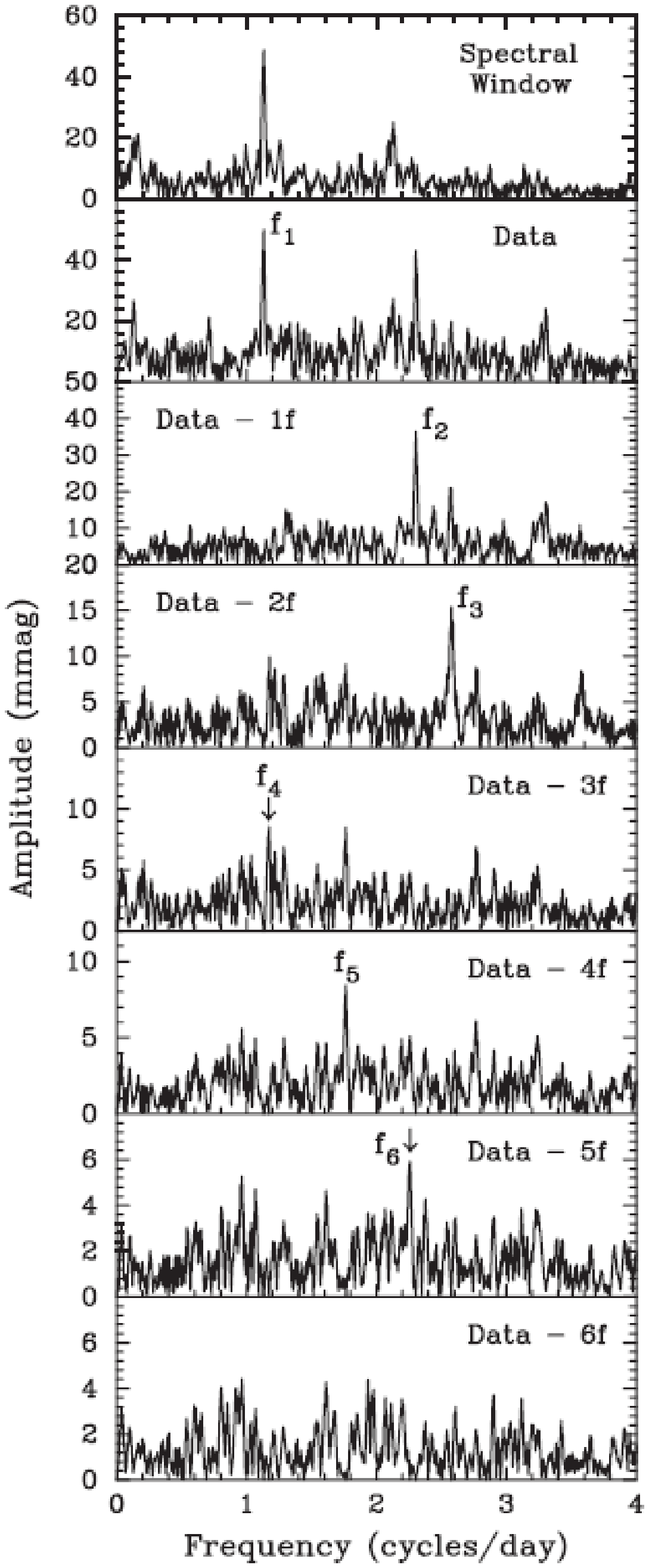}
\includegraphics[height=5in,width=2.2in,angle=0]{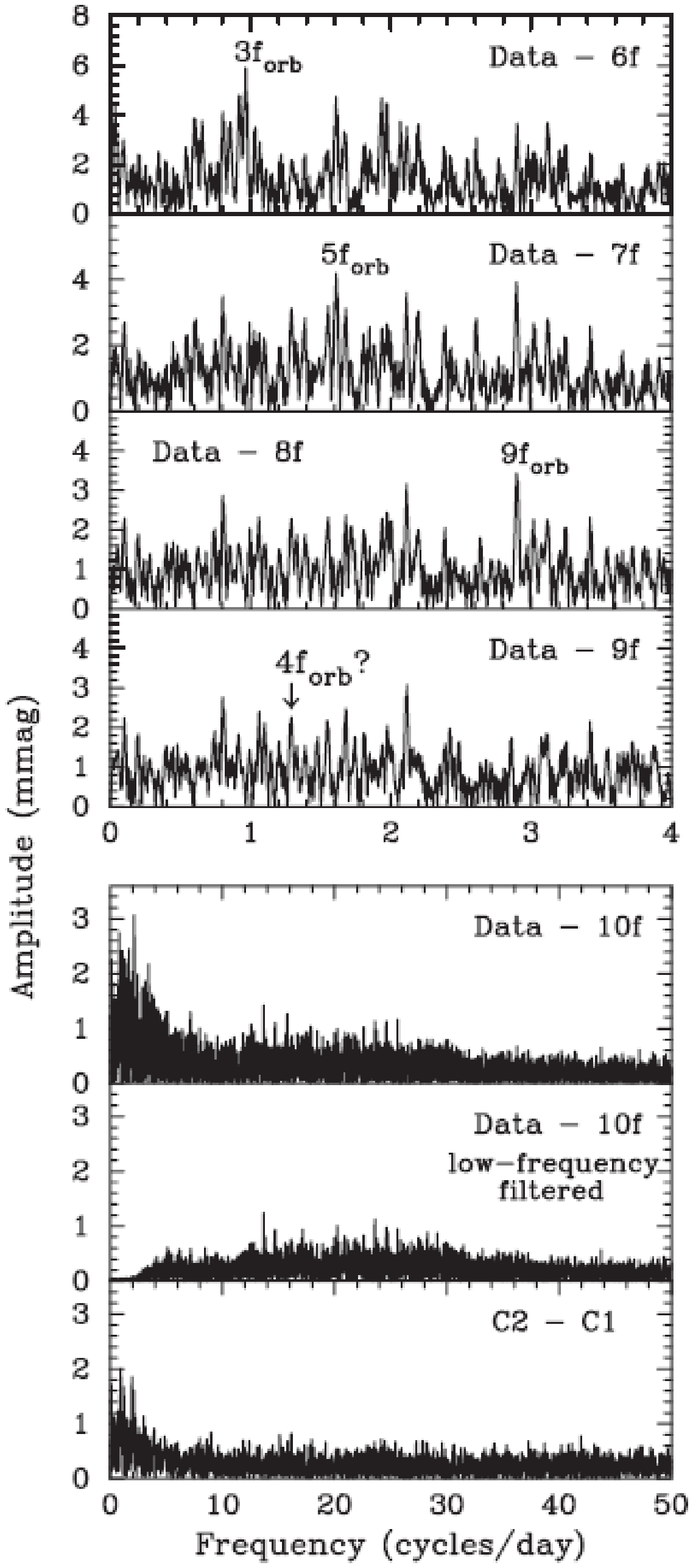}
\caption{Left: Spectral window and amplitude spectra of the B filter time-series
photometry of HD\,209295 after subsequent stages of prewhitening.  Top Right:
amplitude spectra of the combined B and scaled V filter data, after prewhitening
with the 6-frequency solution shown in the left panel. Several harmonics of the
orbital frequency are found. Bottom Right: amplitude spectra of the differential
magnitudes of the comparison stars. Figure reproduced from Handler et al.\
(2002) with permission from MNRAS.}
\label{gerald}
\end{figure}

\subsection{Tidally-excited oscillations}

\begin{figure}
\includegraphics[height=5in,width=5.3in,angle=0]{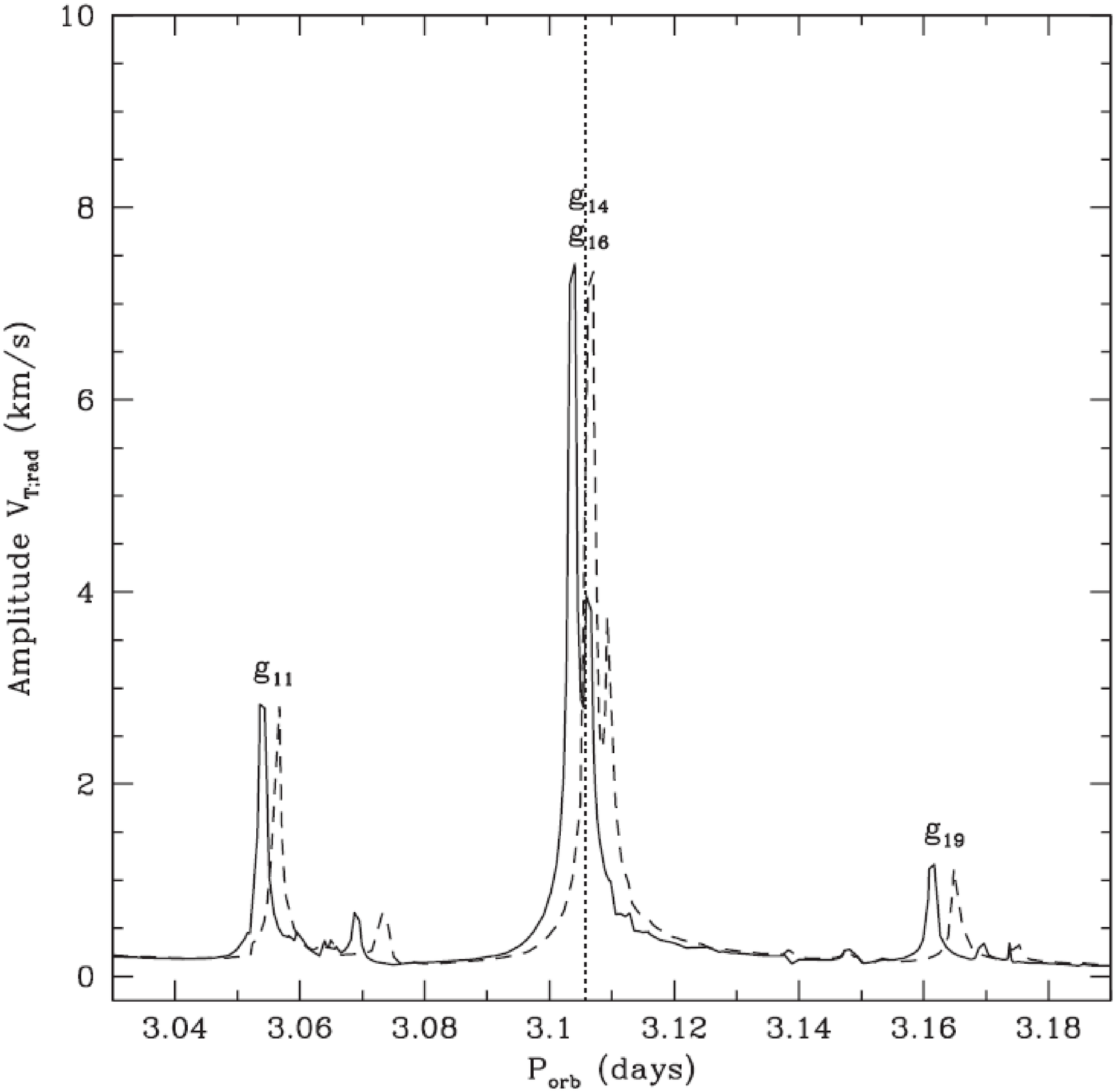}
\caption{Predicted amplitudes of tidally induced radial velocity variations
  in a model for HD\,209295, for two slightly different values of the rotation
  frequency (1.852\,d$^{-1}$: full line; 1.854\,d$^{-1}$: dashed line). The
  orbital frequency of the star is indicated by the dotted vertical line.
  Figure reproduced from Handler et al.\ (2002) with permission from MNRAS.}
\label{tides}
\end{figure}

There are at present only two cases known that meet the requirement of having
tidally-excited oscillation modes. It concerns the slowly pulsating B star
HD\,177863 (B8V), for which De Cat et al.\ (2000) found an oscillation mode
whose pulsation frequency is an exact multiple of 10.00 times the orbital
frequency.  The orbit of the star is very eccentric with $e=0.77$ which provides
a good situation to achieve a resonance between an $\ell =2$ mode with a period
of 1 day and the orbit of 10 days. Willems \& Aerts (2002) made computations
based on tidal oscillation theory for this star and indeed came up with the
possibility that it is undergoing a non-linear resonantly excited $\ell =2$
oscillation mode.  This star has only one confirmed oscillation so far, such
that seismic modelling is not yet possible since this requires at least two
well-identified modes.

By far the most interesting case of tidally-induced oscillations was found by
Handler et al.\ (2002). They discovered the star HD\,209295 (A9/F0V) to be a
binary with $P_{\rm orb}=3.11$d and $e=0.352$ (see Fig.\,\ref{orbit}) exhibiting
one $\delta\,$Sct-type pressure mode and nine $\gamma\,$Dor-type gravity modes,
of which five have an oscillation frequency which is an exact multiple of the
orbital frequency.  The frequency spectra of the star, after subsequent stages
of prewhitening, are shown in Fig.\,\ref{gerald}.  The authors also predicted
frequencies of tidally-excited modes for appropriate fundamental parameters of
the primary and found several such gravity modes to agree with the observed
ones, after having corrected the latter for the surface rotation of the star
(see Fig.\,\ref{tides}). The unfortunate situation of not being able to
identify the modes occurs again for this star, such that seismic tuning of its
structure was not achieved so far.

\section{Conclusions and outlook}

We recall that the requirements for successful seismic modelling of a star are
stringent. We need accurate frequency values, reliable identification of the
spherical wavenumbers $(\ell,m)$ and accurate fundamental parameters with a
precision better than typically 10\% before being able to start the modelling
process. These requirements demand long-term high signal-to-noise and
time-resolved spectroscopy and multicolour photometry with a duty cycle above,
say, 50\%. It is important that the data cover the overall beat-period of the
oscillations, ranging from a few days for compact oscillators up to several
months for main-sequence gravity-mode oscillators. These requirements have been
met recently for a few bright single stars, with impressive improvement for
their interior structure modelling, but not yet for a pulsating star in a close
binary. For those candidates that came close to meeting these requirements, the
problem of mode identification occurred and prevented seismic tuning.

We come to the conclusion that the potential of seismic modelling of close
binaries is {\it extremely good}, particularly for eclipsing binaries. At the
same time, its application is {\it extremely demanding\/} from an observational
point of view. It is clear that efficient progress in this field can only be
achieved from coordinated multisite multitechnique observing campaigns,
preferrably in combination with uninterrupted space photometry. I strongly
encourage the binary and asteroseismology communities to collaborate and take up
this challenging project.

\newpage

\begin{acknowledgments}
The author is much indebted to the Research Council of the Catholic University
of Leuven for significant support during the past years under grant
GOA/2003/04. She is also grateful to the organisers of this meeting for the
opportunity to present this work and to Petr Harmanec for inspiring discussions
and wise lessons on binary stars.
\end{acknowledgments}

\begin{discussion}

\discuss{R.E.\ Wilson}{Have you looked into the time scale for orbital changes
caused by dissipation of orbital energy by excitation of pulsations?}

\discuss{C.\ Aerts}{I haven't, but other peoples did. In any case, such energy
dissipation happens on an evolutionary timescale. This is very different from
apsidal motion, with typical timescales between 10 to 100 years (see talk by
Gimenez). The computation of timescales of energy dissipation through
oscillations (i.e.\ in the case where there is no mass transfer) is difficult,
because it is very dependent on the assumed initial conditions, in particular on
the rotation velocity. Analytical computations of energy loss are availabe in,
e.g., Willems et al.\ 2003, A\&A, 397, 973. Numerical computations were made by,
e.g., Witte \& Savonije 2002, A\&A, 386, 222. All these works, however, predict
too long timescales if we compare them with those derived from data of high-mass
X-ray binaries (the only observed cases available). This might be due to the
much stronger radiative damping during resonances than anticipated so
far. Besides the orbital energy loss, one also has to keep in mind that the
oscillations probably imply significant angular momentum loss of the components
through non-rigid internal rotation.}

\discuss{P.\ Lampens}{How much time was needed to reach the significant new
  insight for single stars from the oscillations that you discussed (non-rigid
  internal rotation and estimate of core overshoot), given that the combination
  of pulsation and binarity is even more demanding?}

\discuss{C.\ Aerts}{We managed to derive the non-rigid internal rotation in two
  main-sequence B stars so far. These stars have multiple oscillation periods of
  the order of several hours. One star was monitored with one and the same
  instrument from a single site during 21 years! The other one was monitored
  during a well-coordinated multisite photometric and spectroscopic campaign
  lasting 5 months and involving about 50 observers (see references in the
  paper). It is therefore clear that asteroseismic inference requires long-term
  monitoring. I think that the binarity does not impose the necessity of even
  longer runs (at least not for close binaries), because the most stringent
  demand is the frequency accuracy and the mode identification, while the
  orbital determination will result naturally and efficiently from a multsite
  spectroscopic effort.}

\end{discussion}

\end{document}